\documentclass[prb,twocolumn,amsmath]{revtex4}
\input {epsf.sty}
\usepackage{epsfig}
\usepackage{amsmath,amsthm,amssymb,graphicx}
\begin{document}
\title{Strong Coupling Corrections to the Ginzburg-Landau Theory of Superfluid $^{3}$He}
\author{H. Choi, J.P. Davis, J. Pollanen, T.M. Haard and W.P. Halperin}
\affiliation{Department of Physics and Astronomy,\\
          Northwestern University, Evanston, Illinois 60208}

\date{Version \today}

\begin{abstract}

In the Ginzburg-Landau theory of superfluid
$^{3}$He, the free energy is expressed
as an expansion of invariants of a complex order parameter.
Strong coupling
effects, which increase with increasing pressure, are embodied in the
set of coefficients of these
order parameter invariants\cite{Leg75,Thu87}.  Experiments can be
used to determine four independent
combinations of the coefficients of the five fourth order invariants. This leaves the phenomenological description of the 
thermodynamics near
$T_{c}$ incomplete. Theoretical understanding of these
coefficients is also  quite limited.
We analyze our measurements of the magnetic susceptibility and the NMR
frequency shift in the $B$-phase which
refine the four experimental inputs to the phenomenological theory.
We propose a model based on
existing experiments, combined with calculations by Sauls and
Serene\cite{Sau81} of the pressure dependence of
these coefficients, in order to determine all five fourth order
terms.  This model
leads us to a better understanding of the thermodynamics of superfluid
$^{3}$He in its various states. We discuss the surface tension of
bulk superfluid $^{3}$He and
predictions for
novel states of the superfluid such as those that are stabilized by
elastic scattering of quasiparticles from a
highly porous silica aerogel.
\end{abstract}

\maketitle

\vspace{11pt}
\section{INTRODUCTION}

The Ginzburg-Landau (GL) formulation gives a phenomenological
representation of
the free energy of superfluid $^{3}$He as an expansion in terms of the order
parameter\cite{Leg75,Thu87,Sau81}.  The expansion coefficients 
specify the stability of
various $p$-wave states and their thermodynamics near $T_{c}$. These 
coefficients are
well-defined theoretically for the weak coupling case. However, 
$^{3}$He is not a weak
coupling superfluid as is clear from  its phase diagram where  there 
is a region of
$A$-phase at high pressures.  This is in contrast to the weak 
coupling limit for which the
$B$-phase is always stable. The strong coupling correction  to 
the pair interaction
is responsible  for the $A$-phase, an effect
of spin and density fluctuations proportional to
$T_{c}/T_{F}$\cite{Rai76}. Calculations\cite{Sau81}
   cannot
account quantitatively for the strong coupling corrections  and so the
coefficients must be
determined empirically. Five of these
parameters are coefficients of the fourth order
invariants of the order
parameter in  the GL free energy.  These are called the
$\beta$-parameters,
$\beta_{i}$'s where
$i =$ 1, ..., 5.  Unfortunately,
there are not enough
independent sets of experiments to determine all the parameters and so the
phenomenological description of superfluid $^{3}$He is under 
determined. This hampers our
ability to predict  stability for novel superfluid $p$-wave states, 
such as those
that might be favored by elastic scattering from high porosity silica aerogel.

In this paper, we present four combinations of $\beta_{i}$'s which we 
determine from
measurements and we describe a
model which resolves the ambiguity in identifying all five of them
independently. The
coefficient of a field dependent term in GL theory,
$g_{z}$, plays an important role in determining  more accurate combinations
of the $\beta_{i}$ than have been previously reported. Our NMR
measurements of the
susceptibility\cite{Haa01} show that
$g_{z}$ is close  to its weak coupling value at all pressures. This
allows us to
interpret our high resolution measurements of the  NMR frequency shift in the
$B$-phase\cite{Kyc94,Kyc97} and to obtain accurate $\beta$-parameter
combinations.

Our model for determining the five $\beta_{i}$'s is motivated by the
calculations of Sauls
and Serene\cite{Sau81}. We note that the calculations, although only
qualitatively
consistent with the existing experiments, nonetheless can accurately
account for
their pressure dependence.  Furthermore, we note that the
calculations indicate that one of the
$\beta$-parameters, $\beta_{1}$, is close to its weak coupling
value at all
pressures. Motivated by these observations and the fact that the experimentally
known  combinations of the $\beta_{i}$'s approach their weak coupling
values at zero
pressure to within 5\%, we make the  following two assumptions: First, the
$\beta$-parameters are, on average, close to their weak coupling
values at zero
pressure and we use this  criterion to select $\beta_{1}$ at zero
pressure. Second,
we take their pressure dependences from the theory which seems to accurately
represent this aspect of the known
$\beta$-parameter combinations. These assumptions are sufficient to
constitute a
model to determine the full suite of
$\beta$-parameters. With this information we can calculate the surface
tension  between $A$- and
$B$-phases in bulk superfluid $^{3}$He and compare with experiment.
We can also
calculate the stability of the  axi-planar state in bulk superfluid
$^{3}$He as a
function of pressure and we can evaluate predictions for  anisotropic $p$-wave
states that are robust in the presence of elastic scattering from
silica aerogel.

\section{GL Theory for $^{3}$He}  A phenomenological macroscopic
description of
phase transitions is given by the GL theory, in which the free energy is
expressed as
an expansion of the order parameter. In the case of superfluid
$^{3}$He, the order  parameter\cite{Thu87,Vol90}, $A$, is a complex
$3\times3$ matrix and the free energy of the system can be expressed as,
\begin{eqnarray} F &=& -\alpha \mathrm{Tr}(AA^{\dagger})+
g_{z}H_{\mu}(AA^{\dagger})_{\mu\nu}H_{\nu} +
\beta_{1}|\mathrm{Tr}(AA^{T})|^{2} \nonumber \\ && +
\beta_{2}[\mathrm{Tr}(AA^{\dagger})]^{2}
+\beta_{3}\mathrm{Tr}(AA^{T}(AA^{T})^{*}) \nonumber \\ && +
\beta_{4}\mathrm{Tr}((AA^{\dagger})^{2})
+\beta_{5}\mathrm{Tr}(AA^{\dagger}(AA^{\dagger})^{*}).
\label{GL_free_energy}
\end{eqnarray}
Here the dipole energy term is neglected. The magnetic field components
are
$H_{\mu}$, and $A^{\dagger}$ and
$A^{T}$ are the Hermitian conjugate and transpose of $A$. The
structure of the order parameter admits five fourth order invariants
each of which has a corresponding coefficient,
$\beta_{i}$. At the second order thermodynamic transition to superfluidity,
$T_{c}$, all
$p$-wave superfluid states are equally probable, but their stability 
below $T_{c}$
depends on the $\beta_{i}$.  In the weak-coupling limit the free 
energy coefficients
are,
\begin{eqnarray} &\alpha={N(0)\over 3} \left({T\over T_{c}}-1\right),\\
&{\beta_{i}\over\beta_{0}} = (-1, 2, 2, 2, -2),  i =1, ..., 5,\\
&\beta_{0}={7\zeta (3)\over 120 \pi^{2}} {N(0) \over (k_{B}T_{c})^{2}},\\
&g_{z}={7\zeta(3)\over48\pi^{2}}N(0) \left( {\gamma_{0}\hbar \over
(1+F_{0}^{a})k_{B}T_{c}} \right)^{2},
\end{eqnarray}
where the normal density of states at the Fermi energy is  $N(0)$, the
gyromagnetic ratio for $^{3}$He is $\gamma_{0}$, $k_{B}$ is the  Boltzmann
constant, $F_{0}^{a}$ is a Fermi liquid parameter determined from the
magnetization measurement\cite{Vol90}  and
$\zeta(x)$ is the Riemann zeta function. However, $^{3}$He is not a weak
coupling superfluid and strong  coupling effects increase with pressure.  The
strong coupling corrections for $\alpha$ are negligible\cite{Sau81} but they
have a significant effect on the $\beta_{i}$'s, and might also 
contribute to $g_{z}$.
Calculations of strong coupling corrections have been performed for  model
potentials\cite{Lev79, Sau81}; those of Sauls and Serene\cite{Sau81} 
being the most
complete and the ones we will refer to in this work.

\section{Experiments}

There are seven free energy coefficients which must be determined from
experiment. The difficulty lies  in the fact
that there is insufficient experimental input to constrain this 
phenomenological
description of superfluid $^{3}$He.  Among the seven coefficients, 
$\alpha$ and $g_{z}$ are
determined without ambiguity. The measurements of the specific heat 
in the normal state,
$C_{N}$ and the  transition
temperature\cite{Gre86}, $T_{c}$, give us $\alpha$. The slope of the
$^{3}$He-$B$ magnetization\cite{Haa01} extrapolated to
$T_{c}$, $dM_{B}/dT|_{T_{c}}$, and
the specific heat jump\cite{Gre86} of $^{3}$He-$B$, $\Delta C_{B}/C_{N}$, are
required for $g_{z}$, for which we have new results presented in this section.
For the remaining five
$\beta_{i}$'s, there are only four independent sets of experiments so 
that only four
combinations of $\beta_{i}$'s can be found in the form of sums. 
These are $\beta_{345}$,
$\beta_{12}$, $\beta_{245}$, and $\beta_{5}$, where we use the 
Mermin-Stare convention,
$\beta_{ij}=\beta_{i}+\beta_{j}$.

First, we will describe the relevant experiments and the logic for determining
these  combinations of the $\beta_{i}$'s.

A). $\beta_{345}$ requires measurements
of the
$^{3}$He-$B$ transverse NMR $g$-shift\cite{Kyc94, Kyc97}, $g$, which  must be
combined with the slope of the $B$-phase longitudinal NMR resonance
frequency\cite{Ran94, Ran96,Kyc97},
$\nu_{B||}^{2}/(1-t)$, in the limit approaching $T_{c}$ as well as 
with measurements of
$\Delta C_{B}/C_{N}$\cite{Gre86} where $t = T/T_{c}$. In order to 
have the value of the
$g$-shift at $T_{c}$ it is helpful to observe that the $B$-phase 
susceptibility and
the $g$-shift are linearly related, facilitating an extrapolation to 
$T_{c}$. The $B$-phase
heat capacity jump is measured only below the polycritical point 
(PCP).  However,
measurement of the specific heat in the $A$-phase along with 
measurements of the latent
heat at the $A$- to $B$-transition allows a thermodynamic 
calculation\cite{Gre86} of the
specific heat jump for the $B$-phase at pressures above the PCP. Consequently, $\Delta C_{B}/C_{N}$ is
experimentally determined at all pressures.

B). From the specific heat jump, $\Delta C_{B}/C_{N}$ and the values 
for  $\beta_{345}$
obtained above we can directly determine  $\beta_{12}$.

C). From the specific heat jump, $\Delta C_{A}/C_{N}$ we can directly determine
$\beta_{245}$, but only for pressures greater than the PCP where this 
jump can be
measured.  Below the PCP $\beta_{245}$ is found from the quadratic 
magnetic field
suppression of the first order
$^{3}$He $A$- to $B$-transition\cite{Tan91}, $g(\beta)$, along with 
the values of
$\beta_{12}$ and $\beta_{345}$ that have been obtained above in A) and B).

D). Finally, we can fix $\beta_{5}$ uniquely by the asymmetry ratio, 
$r$, of the linear
field  dependent splitting of the  $A_{1}$ to $A_{2}$ 
transitions\cite{Isr84} in high
magnetic field combined with $\beta_{245}$.

In summary, four independent combinations of
experiments gives us four constraints on the $\beta_{i}$'s, which is insufficient to identify all five of them. In principle, measurement of the surface tension at
the $^{3}$He
$A$-$B$ interface could provide us with a fifth independent
combination\cite{Thu91,Osh77,Bar04} of $\beta_{i}$'s. However, the  surface
tension vanishes near $T_{c}$ due to the degeneracy of the free energy at
$T_{c}$ of $A$- and $B$-phases.  For this reason it is not possible to  obtain
sufficiently high resolution measurements of the surface tension  to provide
useful characterization of strong coupling effects in the
Ginzburg-Landau limit. In the following, we will discuss in more 
detail the experimental
determination of strong coupling and its effects on the $\beta_{i}$'s.

The coefficient for the field coupling term,
$g_{z}$, is determined by
measuring the slope of the magnetization of $^{3}$He-$B$ in the limit
approaching $T_{c}$,
\begin{equation}
\hat g_{z} \equiv {g_{z} \over g_{z}^{\mathrm wc}}={{dm \over dt}
\over ({dm \over dt})^{\mathrm
wc}}{{\Delta C_{B}^{\mathrm wc}}\over{\Delta C_{B}}},
\end{equation}
where $m=M_{B}/M_{N}$ and $M_{N}$ is the normal state magnetization.
The superscript wc, which we use here and in the following,
indicates the weak
coupling limit.

\begin{table*}
{\small
\begin{tabular*}{0.8\textwidth}{@{\extracolsep{\fill}}|p{0in}cp{0in}||ccccccp{0in}|ccccp{0in}|}
\hline \\*[-12pt]
&$P$& & $g$-shift & $\frac{d\nu^2_{\mathrm{B}\parallel}}{dt}$ & $g(\beta)$ &
$\frac{\Delta C_\mathrm{B}}{C_\mathrm{N}}$ &
$ \frac{\Delta C_\mathrm{A}}{C_\mathrm{N}}$&
$-\frac{(\frac{dT}{dH})_\mathrm{A1}}{(\frac{dT}{dH})_\mathrm{A2}} $&&
$\frac{\beta_{345}}{\beta_0}$ & $ \frac{\beta_{12}}{\beta_0} $ & $
\frac{\beta_{245}}{\beta_0}
$ &
$
\frac{\beta_5}{\beta_0}
$&\\*[0pt]
&bar& &
$\times 10^6$ & $10^{10}$ Hz$^2$ & &
& & & & & & &&\\ \hline\hline
&w.c.& & & & 1 & 1.426 & 1.188 & 1 && 2 & 1 & 2 & -2 & \\ \hline

&0 && 7.31 &1.50 & 1.61 & 1.46 & 1.25 & 0.97 && 2.11 & 0.92 & 1.90 & -1.84&
\\*[-3pt]

&1 && 7.71 & 1.78 & 1.72 &1.50 & 1.29 & 0.99 && 1.86 & 0.97 & 1.84 & -1.82 &\\*[-3pt]

&2 && 8.10 & 2.06 & 1.84 & 1.53 & 1.33 & 1.02 &&1.68 & 0.99 & 1.78 & -1.81 &\\*[-3pt]

&3 && 8.48 & 2.34 & 1.96 & 1.56 & 1.37 & 1.04 && 1.56 & 1.01 &
1.74 & -1.81& \\*[-3pt]

&4 && 8.85 & 2.62 & 2.07 & 1.58 & 1.40 & 1.07 &&
1.47 & 1.01 & 1.70 & -1.81 &\\*[-3pt]

&5 && 9.20 & 2.90 & 2.20 & 1.61 & 1.43 & 1.09 && 1.41 & 1.01 &
1.66 & -1.81 &\\*[-3pt]

&6 && 9.55 & 3.18 & 2.37 & 1.63 &1.46 & 1.12 && 1.36 & 1.01 & 1.63 & -1.82 &\\*[-3pt]

&7 && 9.89 & 3.46 & 2.57 & 1.65 & 1.49 & 1.14 && 1.32 & 1.00 & 1.60 & -1.82&
\\*[-3pt]

&8 && 10.22 & 3.74 & 2.80 &1.67 & 1.51 & 1.16 && 1.29 & 1.00 & 1.57 & -1.83 &\\*[-3pt]

&9 && 10.54 & 4.02 & 3.06 & 1.68 & 1.54 & 1.19 &&
1.26 & 0.99 & 1.55 & -1.84 &\\*[-3pt]

&10 && 10.86 & 4.30 & 3.34 & 1.70 & 1.56 & 1.21 && 1.24 & 0.98
& 1.52 & -1.85 &\\*[-3pt]

&11 && 11.17 & 4.58 & 3.66 & 1.71 & 1.58 & 1.24 && 1.23 & 0.98 & 1.50 &
-1.86 &\\*[-3pt]

&12 && 11.47 & 4.86 & 4.03 & 1.73 & 1.61 & 1.26 && 1.21
& 0.97 & 1.48 & -1.87&
\\*[-3pt]

&13 && 11.77 & 5.14 & 4.51 & 1.74 & 1.63 & 1.29 && 1.20 & 0.96 & 1.46
& -1.88 &\\*[-3pt]

&14 && 12.06 & 5.42 & 5.20 & 1.75 & 1.66 & 1.31 && 1.19 & 0.96 & 1.44 & -1.89& \\*[-3pt]

&15 && 12.36 & 5.70 & 6.21 & 1.77 & 1.68 & 1.34 && 1.18 & 0.95 & 1.41 & -1.89&
\\*[-3pt]

&16 && 12.64 & 5.98 & 7.70 & 1.78 & 1.71 & 1.36 && 1.17 & 0.95 & 1.39 & -1.90 &\\*[-3pt]

&17 && 12.93 & 6.26 & 9.81 & 1.79 & 1.73 &1.39 && 1.15 & 0.94 & 1.37 & -1.90& \\*[-3pt] 

&18 && 13.22 & 6.54 & 12.71 & 1.80 & 1.76 & 1.41 && 1.14
& 0.94 & 1.35 & -1.91& \\*[-3pt] 

&19 && 13.50 & 6.82 & 16.53 & 1.81 & 1.78 & 1.44 && 1.13 & 0.93 &
1.34 & -1.92 &\\*[-3pt] 

&20 && 13.79 & 7.10 & 21.30 & 1.82 & 1.80 & 1.46&
& 1.12 & 0.93 & 1.32 &
-1.93 &\\*[-3pt]

& 21& & 14.08 & 7.38 &  & 1.83 & 1.83 & 1.49 && 1.10 &
0.93 & 1.30 & -1.93&
\\*[-3pt] 

&22 && 14.36 & 7.66 &  & 1.84 & 1.85 & 1.51 && 1.09 & 0.92 & 1.28 &
-1.94 &\\*[-3pt] 

&23 && 14.65 & 7.94 &  & 1.86 & 1.87 & 1.54 && 1.08 & 0.92 & 1.27 & -1.95&
\\*[-3pt] 

&24 && 14.95 & 8.22 &  & 1.87 & 1.90 & 1.56 && 1.06 & 0.92 & 1.25 & -1.96 &\\*[-3pt] 

&25 && 15.25 & 8.50 &  & 1.88 & 1.92 & 1.58 &&
1.05 & 0.92 & 1.24 & -1.97 &\\*[-3pt] 

&26 && 15.55 & 8.78 &  & 1.89 & 1.94 & 1.61 && 1.03 & 0.91 &
1.23 & -1.97 &\\*[-3pt] 

&27 && 15.85 & 9.06 &  & 1.90 & 1.96 & 1.63 &&
1.02 & 0.91 & 1.21 & -1.98 &\\*[-3pt] 

&28 && 16.17 & 9.34 &  & 1.91 & 1.98 & 1.66 && 1.00 & 0.91 &
1.20 & -1.99 &\\*[-3pt]

&29 && 16.49 & 9.62 &  & 1.92 & 2.00 & 1.68 && 0.99 & 0.91 & 1.19 & -2.00 & \\*[-3pt] 

&30& & 16.81 & 9.90 &  & 1.93 & 2.02 & 1.71 && 0.97 & 0.91 & 1.18 & -2.01 &\\*[-3pt]

&31& &  &  &  &   & 2.04 & 1.73& &  & & 1.16 & -2.02 &\\*[-3pt] 

&32& &  &  &  &  & 2.07 & 1.76& &  & &  1.15 &
-2.02 &\\*[-3pt] 

&33 &&  &  &  &  & 2.09 & 1.78& &  & &  1.14 & -2.03 &\\*[-3pt] 

&34& &  &  &  &  & 2.12& 1.81 &&  & &  1.12 & -2.03 & \\ \hline
\end{tabular*} }
\caption{Ginzburg-Landau $\beta$-parameters and the experimental
quantities from which they are
derived. The NMR $B$-phase $g$-shift is a fit to data from
Kycia\cite{Kyc97} given by Haard\cite{Haa01}.  The NMR $B$-phase
longitudinal
resonance  was measured by Rand\cite{Ran94,Ran96} for which a
smoothed fit is given by Haard\cite{Haa01}.   The coefficient of
quadratic magnetic field suppression of the $B$-phase was measured by
Tang {\it et al.}\cite{Tan91}. The $B$-phase heat capacity
jump was taken fromß Greywall\cite{Gre86} and the asymmetry ratio of
the linear field dependent splitting of the  $A_{1}$ to $A_{2}$
transitions was reported by Israelson {\it et al.}\cite{Isr84}. 
Extension of the
measuerments of the $A$-phase heat capacity jump to pressures lower 
than the PCP requires
a calculation based on the measured quadratic suppression of the $A$- 
to $B$-transition as
described in the text.}
\label{beta_combo}
\end{table*}

\begin{figure}[t]
\centerline{\includegraphics[width=3.4in]{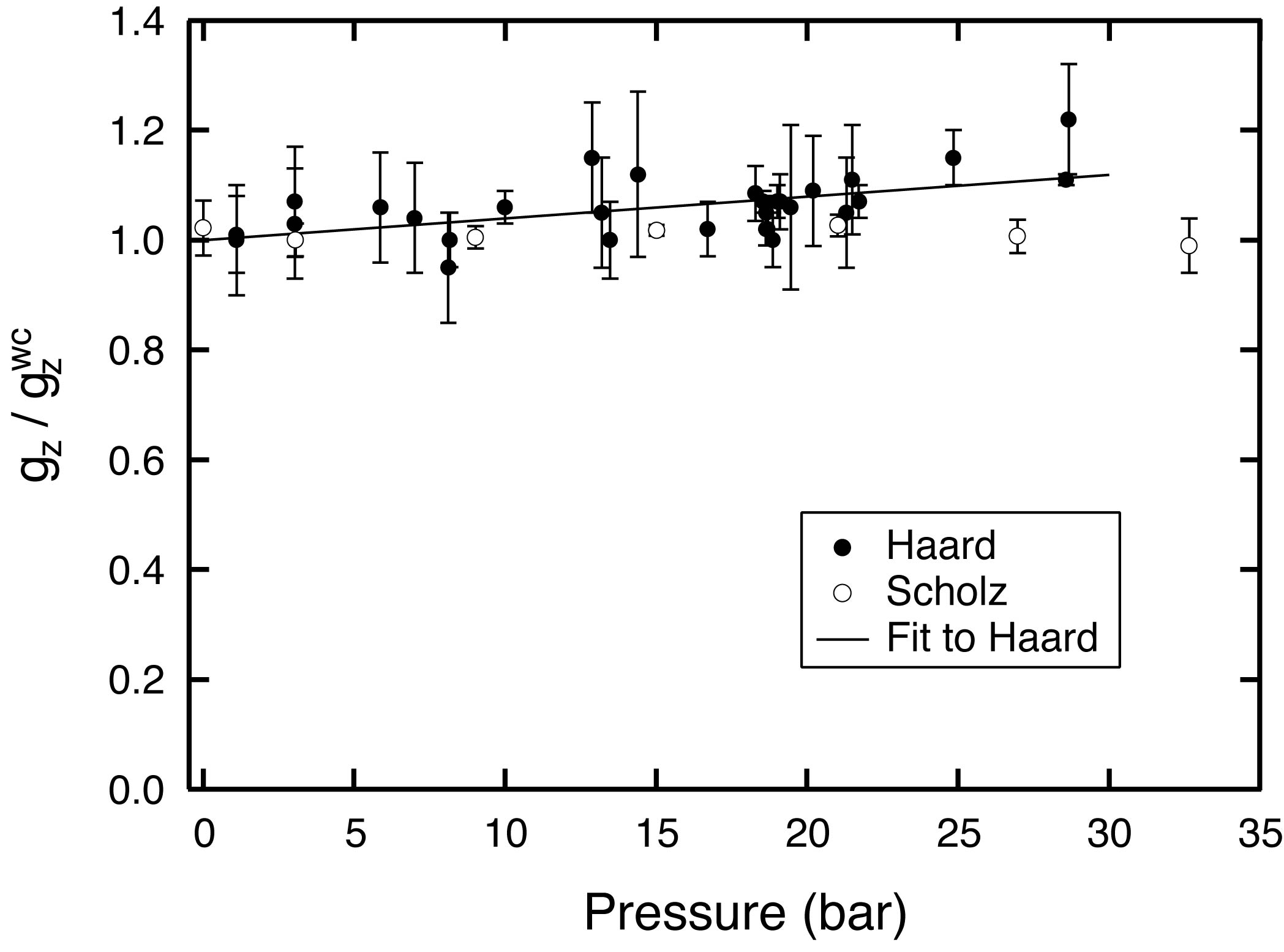}}
\begin{minipage}{0.93\hsize}
\caption {$\hat g_{z}$ obtained from magnetization measurements by
NMR. Closed circles are the
measurements by Haard\cite{Haa01} and open circles by
Scholz\cite{Hoy81, Sch81}. The results from
both measurements are consistent and give approximately unity for $\hat g_{z}$}
\end{minipage}
\end{figure}

Magnetization measurements of superfluid $^{3}$He have been of great
interest since its discovery. Two different techniques - NMR based  dynamic
measurements\cite{Cor75, Aho75, Aho76, Hoy81, Haa01} and SQUID based
static measurements\cite{Pau73,  Pau74, Osh74, Hah95} - have been
performed over the past thirty years. Historically, there has been a
discrepancy between these two techniques\cite{Web77,  Sag77} the origin of which has not been established.  Nonetheless, more  recent
experiments\cite{Haa01, Hah98} bring the results closer together. Haard
measured the magnetization using high resolution NMR\cite{Haa01}. A
careful analysis of this and other  measurements\cite{Sch81,Hah98} reveals
that the discrepancy appears to be negligible near the transition
temperature $T_{c}$. Using Eq. 6, Haard\cite{Haa01} determined  $g_{z}$
from NMR and found  the results presented in Fig. 1, where $g_{z}$ is close
to its weak coupling value, i.e. $\hat g_{z}=1$.  From Haard's measurement,
the  deviation from weak coupling appears to grow slightly with pressure. From
analysis\cite{Haa01} of the more accurate work of Scholz {\it et 
al.}\cite{Hoy81, Sch81}
it appears that $g_{z}$ is pressure independent.  The difference 
between the data sets is
likely due to the wider  range of extrapolation in the
$B$-phase toward $T_{c}$ that is required to determine $g_{z}$ at elevated
fields in the  case for Haard's measurement. Hahn {\it et
al.}\cite{Hah98} came to the same conclusion, $\hat g_{z} = 1$, based  on
their SQUID measurements, and so we will take $g_{z}$ to have its weak
coupling value at all pressures. Having established $g_{z}$, $\beta_{345}$
can be calculated from the  NMR $g$-shift\cite{Kyc94, Kyc97,Haa01,Moo93} of
the transverse NMR frequency in $^{3}$He-$B$  which has the following
relationship\cite{Gre76} with
$\hat g_{z}$ and $\beta_{345}$:
\begin{equation}
\frac{\beta_{345}}{\hat g_{z}}=
\frac{\beta_{345}^{\mathrm wc}}{(1+F_{0}^{a})^{2}} \left(
\frac{C_{N}}{\Delta C_{B}}\right) \frac{\nu_{B \parallel}^{2}}{1-t}
\left( \frac{\hbar}{2 \pi k_{B} T_{c}} \right)^{2} \frac{1}{g}.
\label{Eq2}
\end{equation}

In earlier
reports\cite{Kyc94} of the $g$-shift, the analysis to obtain
$\beta_{345}$  estimated $g_{z}$ incorrectly. The
values in Table I for the $g$-shift and the $B$-phase longitudinal
resonance frequency are smoothed values\cite{Haa01} from
a large number of experiments\cite{Kyc97}, significantly more than 
what was originally
reported by Kycia {\em et al}.\cite{Kyc94}.  Greywall\cite{Gre86} has measured the specific
heat of $^{3}$He-$A$ and $B$. The specific heat jump at
$T_{c}$, for these two phases, is related to
$\beta_{A}$ and $\beta_{B}$ through:
\begin{eqnarray} &{\Delta C_{A}}=
\frac{\alpha'^{2}}{2\beta_{A}}, \beta_{A} \equiv \beta_{245} \\ &{\Delta
C_{B}}= \frac{\alpha'^{2}}{2\beta_{B}}, \beta_{B}
\equiv \beta_{12} +
\frac{1}{3}\beta_{345},
\label{Eq4}
\end{eqnarray}
where $\alpha\,' \equiv d\alpha/dT$.
At pressures less than the PCP, the magnetic
suppression\cite{Tan91}, $g(\beta)$, of the $AB$ transition temperature, $T_{AB}$, is used to obtain $\beta_{245}$ through:
\begin{equation} g(\beta) =
-{{\sqrt{1+(\beta_{B}/\beta_{A}-1)(1+{{2}\over{1-\beta_{12}/\beta_{B}}})}+1}\over{\beta_{B}/\beta_{A}-1}}.
\end{equation}
Here $g(\beta)$ is defined by,
\begin{equation} 1- \frac{T_{AB}}{T_{c}} \equiv g(\beta)\left(B \over
B_{0}\right)^2 +  {\mathcal O}\left( \left( B \over B_{0}\right)^4 \right),
\end{equation}
where $B$ is the applied magnetic field and $B_{0}^2 = N(0)/6g_{z}$. Finally, $\beta_{5}$ can be determined
by measuring the asymmetry ratio\cite{Isr84} of the $A_{1}$-$A_{2}$
splitting, $r$,
\begin{equation} r \equiv  \frac{T_{A1}-T_{c}}{T_{c}-T_{A2}} =
-\frac{\beta_{5}}{\beta_{245}}.
\end{equation}

The four experimentally determined $\beta$-coefficient combinations,
along with the measurements used to obtain them, are tabulated
from 0 to 34 bar in Table I.

\section{Model for Determining $\beta$'s}

As stated earlier, we impose two assumptions to eliminate ambiguity
associated with sorting out all five $\beta_{i}$'s from the four
known combinations of $\beta_{i}$'s determined from the 
experiments described in
the previous section. The assumptions are: 1) the pressure dependence of
$\beta_{1}$ calculated by Sauls and Serene\cite{Sau81} is valid. 2) At zero
pressure, all five $\beta_{i}$'s approach their weak  coupling values, on
the average. The consequences of these assumptions will be discussed in the
following subsections.

\subsection{Comparison with the Calculation}

\begin{figure}[b]
\centerline{\includegraphics[width=3.4in]{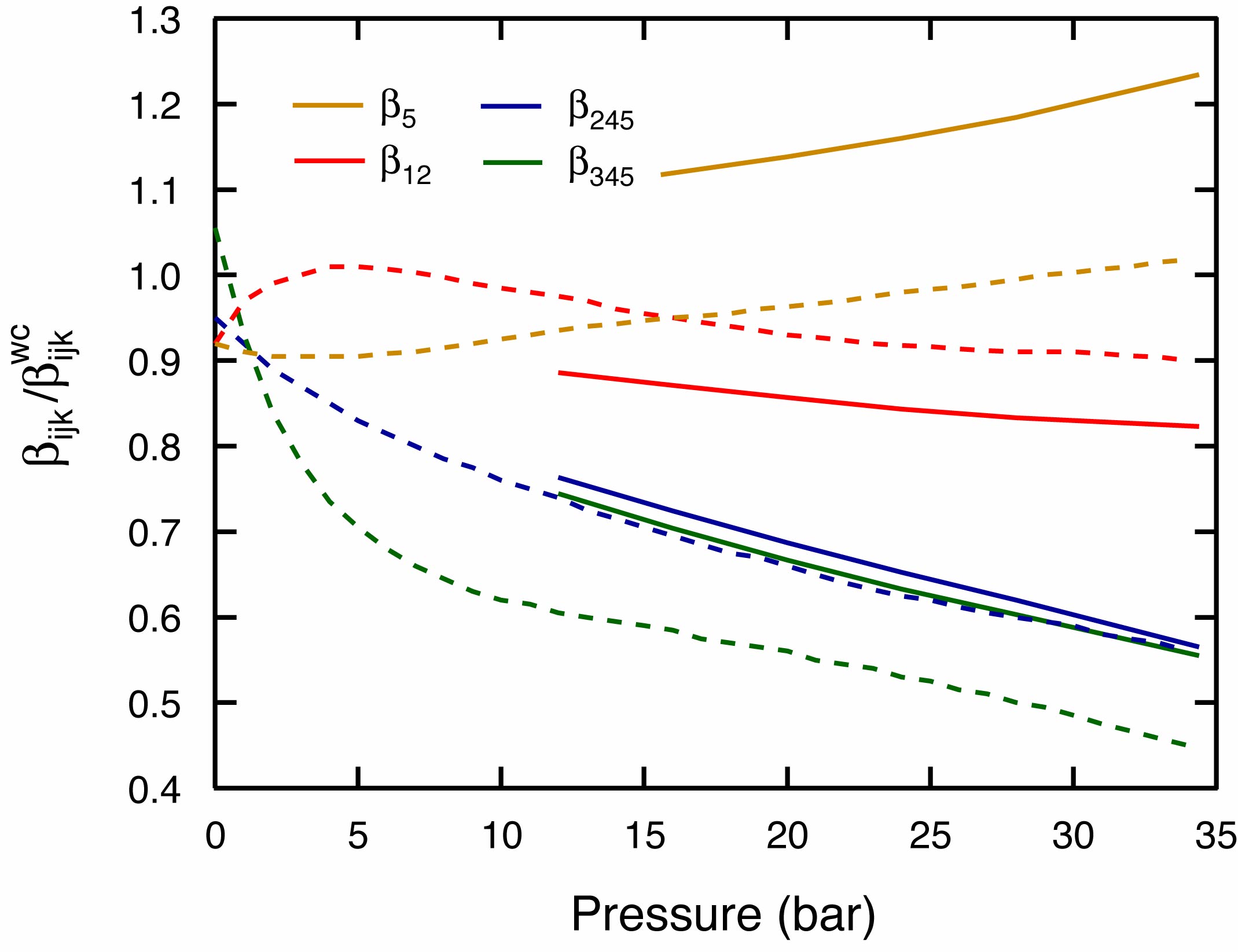}}
\begin{minipage}{0.93\hsize}
\caption {Comparison of four known $\beta$-combinations from the
experiments (dashed lines) and
Sauls and Serene's calculation\cite{Sau81} (solid lines). The
pressure dependences are in good agreement but
the absolute values are not as close.}
\end{minipage}
\end{figure}
\noindent

Sauls and Serene\cite{Sau81} developed a potential scattering model
   to find the strong coupling corrections to the
$\beta$-coefficients in the pressure range of 12 to 34.4 bar. Since we do
not have five experimentally determined
$\beta$-coefficients with which to directly compare to the theory, we
construct from the calculation those four combinations of
$\beta$-coefficients, $\beta_{345}$, $\beta_{12}$,
$\beta_{245}$, and $\beta_{5}$ that are experimentally accessible and
compare these with the measurements  in Fig. 2. First, we note that the
experimental results suggest that  superfluid $^{3}$He is predominantly weak
coupling at zero pressure. Secondly, the pressure dependence
of each combination shows remarkable agreement between experiment and 
theory for $P > 12$
bar, the range where the calculations were performed. It is also 
apparent that the
calculation of the absolute values of the
$\beta_{i}$'s is less reliable than their pressure dependence.  Finally  we
note that, in the calculation, the smallest strong coupling correction
among the $\beta_{i}$'s
    is for $\beta_{1}$. Guided by this information, we will assume that
the pressure dependence of $\beta_{1} (P)$ can be taken from the 
Sauls and Serene
calculation and then we need only determine $\beta_{1}(0)$.

\subsection{Zero Pressure Values of the $\beta_{i}$'s} The pressure dependence
of the $\beta_{i}$'s is insufficient to resolve the
ambiguity associated with the
$\beta$ coefficient combinations. Five independent values of $\beta_{i}$'s
at a given pressure are  required along with the pressure dependence for $\beta_{1}$. The calculations indicate that strong
coupling corrections are smallest for $\beta_{1}(P)$ and from experiment
we see that the measurable combinations deviate from their weak coupling
values by less than  5\% at zero pressure.  On this basis one possibility
would be to simply choose
$\beta_{1}(0)/\beta_{1}^{\mathrm{wc}} = 1$, {\it i.e.} to be weak coupling.
Another possibility, the more democratic one, is to choose
$\beta_{1}(0)$
    as a variational parameter and minimize the mean square deviations of
all $\beta$-parameters from their weak coupling values  at zero pressure
subject to the constraints imposed by the four different combinations  that
have been determined experimentally. For the latter method we find
$\beta_{1}(0)/\beta_{1}^{\mathrm{wc}} = 0.97$ which is  essentially
equivalent to the first choice. In the following we  make the latter
choice. We show this process  explicitly in Fig. 3 where we calculate all of
the
$\beta_{i}(0)$'s as a function of $\beta_{1}(0)$ subject to the four
experimental constraints. It is clear that for $\beta_{1}(0)$ near its weak
coupling value,  as emphasized by the circled region, all the others
approach their weak coupling values at zero pressure as well. With this choice for
$\beta_{1}(0)$
    and the pressure dependence of $\beta_{1}$ taken from Sauls and
Serene\cite{Sau81},
$\beta_{1}(P)$ is now uniquely defined  and all the other $\beta_{i}$'s can be
determined. These $\beta_{i}$'s are tabulated in the first five columns of
Table II.

\begin{figure}[b]
\centerline{\includegraphics[width=3.4in]{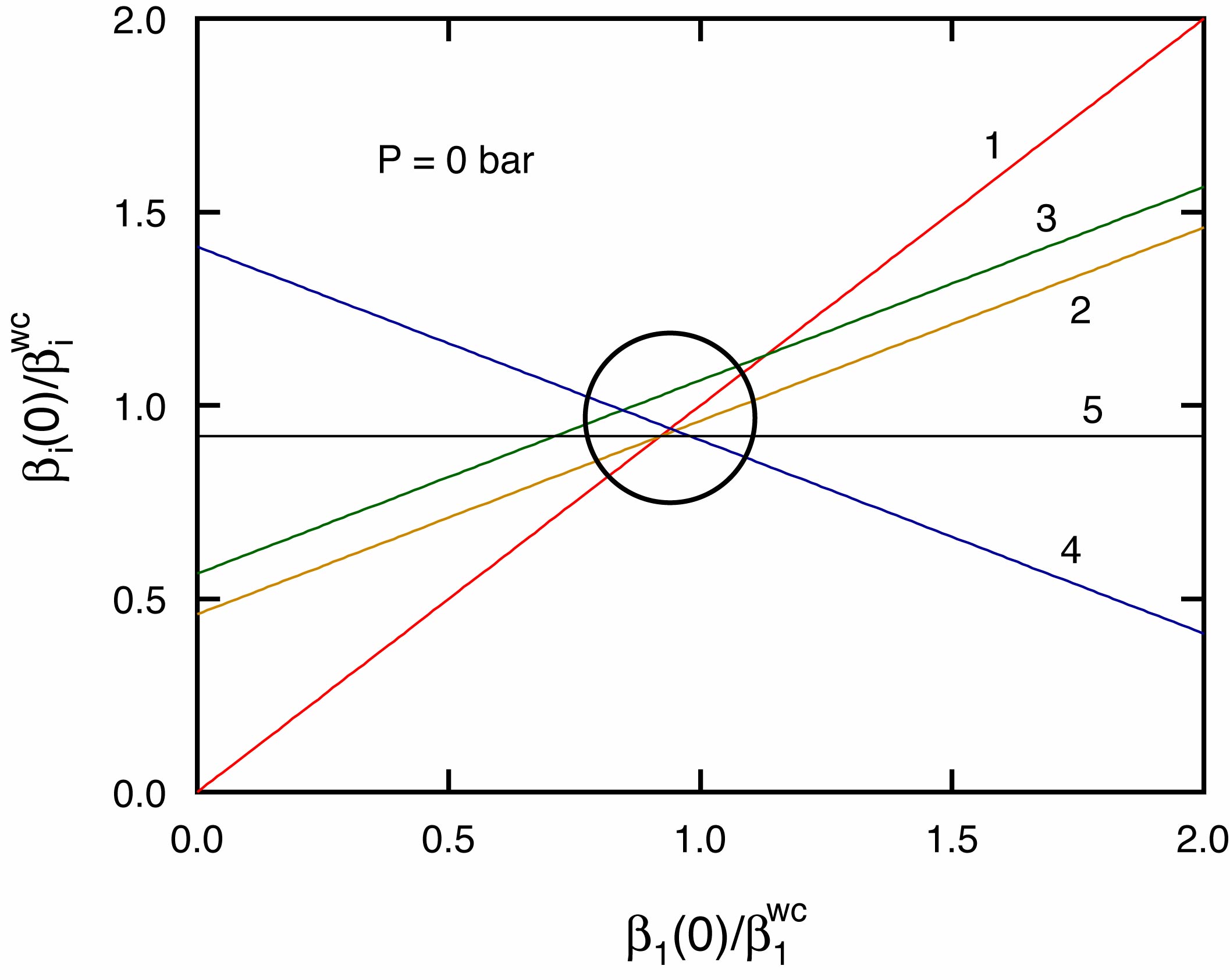}}
\begin{minipage}{0.93\hsize}
\caption {Zero pressure values of the $\beta_{i}$'s parameterized by
$\beta_{1}(0)$. The numbers on each line correspond to the subscript $i$ of
$\beta_{i}$. The $\beta_{i}(0)$'s clearly converge around
$\beta_{i}(0)/\beta_{i}^{wc}=1$, which is an indication that the
$\beta_{i}$'s tend toward their weak-coupling values at low pressure.}
\end{minipage}
\end{figure}
\noindent

\section{Applications}
\subsection{Surface Tension at the $A$-$B$ Interface}

\begin{table*}[t]
{\small
\begin{tabular*}{0.8\textwidth}{@{\extracolsep{\fill}}|p{0in}cp{0in}||cccccp{0in}|cccccp{0in}|}
\hline
&$P$&& \multicolumn{6}{c|}{Pure $^{3}$He} &\multicolumn{6}{c|}{$^{3}$He in
98\% aerogel}\\*[2pt]
\multicolumn{3}{|c||}{(bar)}& $\frac{\beta_1}{\beta_0}$ & $ \frac{\beta_2}{\beta_0} $ &
$\frac{\beta_3}{\beta_0}$ & $\frac{\beta_4}{\beta_0}$ &
$\frac{\beta_5}{\beta_0}$ && $\frac{\beta_1^{a}}{\beta_0^{a}}$ &
$\frac{\beta_2^{a}}{\beta_0^{a}}$ & $\frac{\beta_3^{a}}{\beta_0^{a}}$
&
$\frac{\beta_4^{a}}{\beta_0^{a}}$ & $\frac{\beta_5^{a}}{\beta_0^{a}}$&
\\*[5pt] \hline\hline
&w.c.& & -1 & 2 & 2 & 2 & -2 && -1 & 2 & 2 & 2 & -2&
\\ \hline
&   0 && -0.97 & 1.89 & 2.10 & 1.85 & -1.84 & && &  &  &  & \\*[-3pt]
&   1 && -0.97 & 1.94 & 1.96 &1.72 & -1.82 &  & && &  &  & \\*[-3pt]
&   2 && -0.97 & 1.96 & 1.86 & 1.63 & -1.81 &  & && &  &  & \\*[-3pt]
&   3 && -0.98 & 1.99 & 1.81 & 1.56 & -1.81 &  & && &  &  & \\*[-3pt]
&   4 && -0.98 & 1.99 & 1.76 & 1.52 &-1.81 &  & && &  &  & \\*[-3pt]
&   5 && -0.98 & 1.99 & 1.74 & 1.48 & -1.81 && -0.05 & 0.15 & 0.10 & 0.15
& -0.15&\\*[-3pt]
&   6 && -0.98 & 1.99 & 1.72 & 1.46 & -1.82 && -0.20 & 0.51 & 0.36 & 0.48
& -0.50&\\*[-3pt]
&   7 && -0.98 & 1.98 & 1.70 & 1.44 & -1.82 && -0.28 & 0.72 & 0.53 & 0.66
& -0.70&\\*[-3pt]
&   8 && -0.98 & 1.98 & 1.70 & 1.42 & -1.83 && -0.35 & 0.87 & 0.66 & 0.78
& -0.84&\\*[-3pt]
&   9 && -0.99 & 1.98 & 1.69 & 1.41 & -1.84 && -0.41 & 0.99 & 0.75 & 0.87
& -0.95&\\*[-3pt]
&10 && -0.99 & 1.97 & 1.69 & 1.40 & -1.85 && -0.45 & 1.08 & 0.83 & 0.94
& -1.05&\\*[-3pt]
&11 && -0.99 & 1.97 & 1.70 & 1.39 & -1.86 && -0.49 & 1.15 & 0.90 & 0.99
& -1.12&\\*[-3pt]
&12 && -0.99 & 1.96 & 1.69 & 1.39 & -1.87 && -0.52 & 1.21 & 0.96 & 1.03
& -1.17&\\*[-3pt]
&13 && -0.99 & 1.95 & 1.69 & 1.39 & -1.88 && -0.55 & 1.26 & 1.01 & 1.06
& -1.22&\\*[-3pt]
&14 && -1.00 & 1.95 & 1.70 & 1.38 & -1.89 && -0.58 & 1.30 & 1.05 & 1.09
& -1.27&\\*[-3pt]
&15 && -1.00 & 1.95 & 1.72 & 1.35 & -1.89 && -0.60 & 1.34 & 1.10 & 1.10
& -1.32&\\*[-3pt]
&16 && -1.00 & 1.95 & 1.73 & 1.34 & -1.90 && -0.62 & 1.38 & 1.13 & 1.12
& -1.35&\\*[-3pt]
&17 && -1.00 & 1.94 & 1.72 & 1.33 & -1.90 && -0.64 & 1.40 & 1.16 & 1.13
& -1.38&\\*[-3pt]
&18 && -1.00 & 1.94 & 1.73 & 1.32 & -1.91 && -0.66 & 1.43 & 1.19 & 1.14
& -1.41&\\*[-3pt]
&19 && -1.00 & 1.93 & 1.72 & 1.33 & -1.92 && -0.67 & 1.45 & 1.21 & 1.16
& -1.43&\\*[-3pt]
&20 && -1.01 & 1.94 & 1.74 & 1.31 & -1.93 && -0.69 & 1.48 & 1.24 & 1.16
& -1.47&\\*[-3pt]
&21 && -1.01 & 1.94 & 1.74 & 1.29 & -1.93 && -0.71 & 1.50 & 1.26 & 1.16
& -1.49&\\*[-3pt]
&22 && -1.01 & 1.93 & 1.74 & 1.29 & -1.94 && -0.72 & 1.51 & 1.28 & 1.17
& -1.51&\\*[-3pt]
&23 && -1.01 & 1.93 & 1.74 & 1.29 & -1.95 && -0.73 & 1.53 & 1.30 & 1.18
& -1.53&\\*[-3pt]
&24 && -1.01 & 1.93 & 1.74 & 1.28 & -1.96 && -0.74 & 1.54 & 1.32 & 1.18
& -1.54&\\*[-3pt]
&25 && -1.01 & 1.93 & 1.74 & 1.28 & -1.97 && -0.75 & 1.56 & 1.33 & 1.18
& -1.58&\\*[-3pt]
&26 && -1.02 & 1.93 & 1.73 & 1.27 & -1.97 && -0.76 & 1.57 & 1.34 & 1.18
& -1.60&\\*[-3pt]
&27 && -1.02 & 1.93 & 1.74 & 1.26 & -1.98 && -0.77 & 1.58 & 1.36 & 1.18
& -1.61&\\*[-3pt]
&28 && -1.02 & 1.93 & 1.73 & 1.26 & -1.99 && -0.78 & 1.60 & 1.37 & 1.19
& -1.62&\\*[-3pt]
&29 && -1.02 & 1.93 & 1.73 & 1.26 & -2.00 && -0.78 & 1.61 & 1.38 & 1.19
& -1.63&\\*[-3pt]
&30 && -1.02 & 1.93 & 1.72 & 1.26 & -2.01 && -0.79 & 1.62 & 1.38 & 1.19
& -1.67&\\*[-3pt]
&31 && -1.03 & 1.93 & 1.73 & 1.25 & -2.02 && -0.80 & 1.62 & 1.40 & 1.19
& -1.68&\\*[-3pt]
&32 && -1.03 & 1.93 & 1.73 & 1.25 & -2.02 && -0.81 & 1.63 & 1.40 & 1.19
& -1.68&\\*[-3pt]
&33 && -1.03 & 1.93 & 1.73 & 1.25 & -2.03 && -0.81 & 1.63 & 1.41 & 1.19
& -1.69&\\*[-3pt]
&34 && -1.03 & 1.93 & 1.73 & 1.25 & -2.03 && -0.82 & 1.64 & 1.42 & 1.20 & -1.70&\\
\hline
\end{tabular*} }
\caption{$\beta_{i}$'s for bulk superfluid $^{3}$He, left side, and
superfluid $^{3}$He in 98\% porosity aerogel in the IISM
with $\lambda$= 150 nm and $\xi_{a}$=40 nm, right side.}
\label{beta}
\end{table*}

\begin{figure}[b]
\centerline{\includegraphics[width=3.4in]{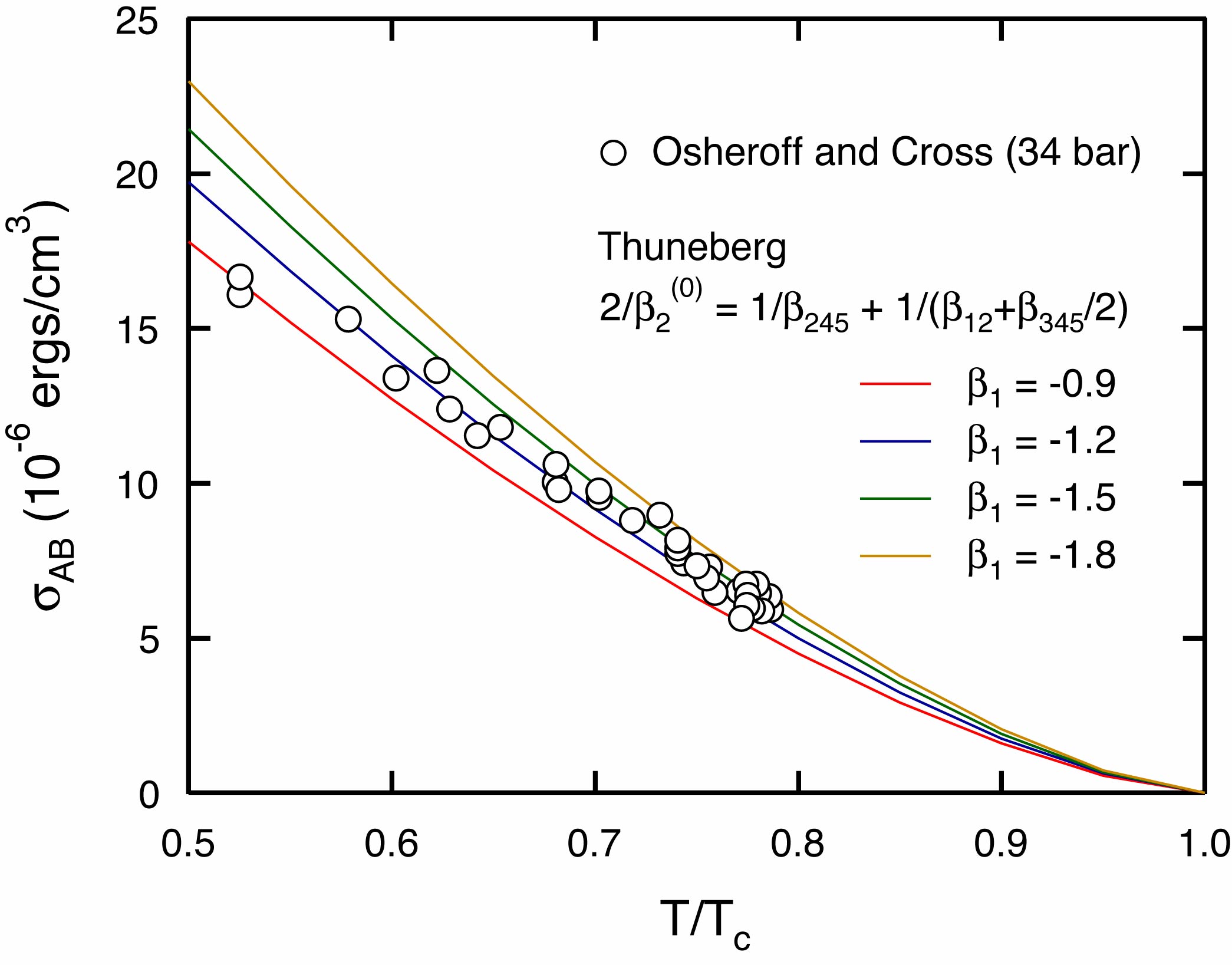}}
\begin{minipage}{0.93\hsize}
\caption {Osheroff and Cross's measurements of surface
tension\cite{Osh77} in comparison to the calculation by
Thuneberg\cite{Thu91} for various $\beta_{1}$ choices and $2/\beta_{2}^{(0)}=\beta_{245}^{-1}+(\beta_{12}+\beta_{345}/2)^{-1}$. The comparison of
the two with the choice of $\beta_{1}/\beta_{1}^{wc}\sim 1$ is consistent,
but the measurement lacks the resolution to conclusively determine the
value of $\beta_{1}$.}
\end{minipage}
\end{figure}
\noindent

With all the $\beta_{i}$'s now determined, one can calculate the
surface tension between the
$A$- and $B$-phases of superfluid $^{3}$He. According to
Thuneberg\cite{Thu91}, the surface free energy of the $A$-$B$ interface is
expressed as
\begin{eqnarray} f_{AB} & = & {\xi(T)\alpha^{2} \over 4\beta_{2}^{(0)}}
\times  \\ \nonumber & &\left[ {I_{1} \over
\sqrt{2\beta_{2}^{(0)}}} + {I_{2} \over 2} \left({4a^{3} \over
\beta_{2}^{(0)}\beta_{3}^{(0)}(\beta_{1}^{(0)}+3\beta_{2}^{(0)})}\right)^{1
\over 4}
\right]
\end{eqnarray} where $\xi(T)$ is the temperature dependent coherence length of $^{3}$He,
\begin{eqnarray}  
I_{1} & = & \left( \begin{array}{ll}
\sqrt{a+c} + {a \over \sqrt{c}}
\mathrm{ln}\left({\sqrt{a+c}+\sqrt{c}\over
\sqrt{a}}\right) & \mbox{if $c>0$}\\
\sqrt{a+c} + {a \over \sqrt{-c}}
\mathrm{arcsin}(\sqrt{-c/a}) & \mbox{if $c<0$,}\\
\end{array} \right.\\
I_{2} & \approx & 1.89 - 1.98\sqrt{\kappa} - 0.31
\kappa \text{ for } \kappa < 1/\sqrt{2}, \\
\kappa^{2} & = & {{\beta_{3}^{(0)}(\beta_{1}^{(0)}+3\beta_{2}^{(0)})}
\over {4\beta_{2}^{(0)}\beta_{34}^{(0)}}}.
\end{eqnarray}
Here $a$ and $c$ are defined as $a=2\beta_{1}+\beta_{3}-\beta_{45}$ and
$c=-(2\beta_{1}+\beta_{345})$. The $\beta_{i}^{(0)}$'s are any set of
$\beta_{i}$'s that satisfy the condition for the surface energy to vanish,
$2\beta_{1}+\beta_{3}=0, \beta_{45}=0$. Weak coupling values of
$\beta_{i}$'s are a subset of the
$\beta_{i}^{(0)}$'s, but the $\beta_{i}^{(0)}$'s need not be limited to
their weak coupling  values. Thuneberg suggested two different values for
$\beta_{2}^{(0)}$,
$2/\beta_{2}^{(0)}=\beta_{245}^{-1}+(\beta_{12}+\beta_{345}/2)^{-1}$ and $\beta_{2}^{(0)}=\beta_{12}+\beta_{345}/2$,  but
kept the relative magnitude of the five $\beta_{i}^{(0)}$'s the same as for
the  weak-coupling case in his original work\cite{Thu91}. We examined
these two choices of $\beta_{2}^{(0)}$.

   From a number of different choices for the $\beta_{1}$ including the values
of the $\beta_{1}$ chosen from Table I, the surface  tension at the melting
curve is calculated. The calculation with our choice of $\beta_{1}$ and the measurements of Osheroff and Cross\cite{Osh77} are in good agreement. An example of the calcuation with $2/\beta_{2}^{(0)}=\beta_{245}^{-1}+(\beta_{12}+\beta_{345}/2)^{-1}$ is shown in Fig. 4.
 The calculation, however, has a number of limitations. One is
that the calculation depends on the choice of
$\beta_{2}^{(0)}$ which is not uniquely defined. The other is that the
experimental results do not have high enough resolution to determine the
$\beta_{i}$'s independently.

\subsection{How Stable Is the Axial State?}

A number of experiments performed to investigate the order parameter of $^{3}$He-$A$ phase have confirmed that the $A$-phase is, in fact, the axial state. This confirmation could be further strengthened by studying the thermodynamic stability of the axial state over other possible equal spin pairing states, such as an axi-planar state; some concern has been raised in the past that an axial state and an axi-planar state may not be easily distinguishable due to their continuously related order parameter structures\cite{Tan91,Gou92}. However, a certain combination of $\beta$-coefficients, namely $\beta_{45}$ can be used to check the relative thermodynamic stability between the two states. If $\beta_{45}$ is negative, the $A$-phase is the axial state and if $\beta_{45}$ is positive, the $A$-phase would be the axi-planar state. By
imposing $\beta_{45}=0$ as the fifth constraint in addition to the four
known  combinations of the $\beta_{i}$'s, a unique set of $\beta_{i}$'s is
obtained which can be used to plot a phase diagram for axial and axi-planar
states with $\beta_{1}$ as a
parameter.  This phase diagram is shown in Fig. 5. We compare our choice
of $\beta_{1}$ from Table I with the phase diagram and this value lies
well within the axial state regime at high pressure as has been commonly
believed and which  a number of experiments independently
confirm\cite{Ran94,Mul94, Ran96}. However, it should be noted that our choice of $\beta_{1}$ indicates that there is a near degeneracy of the axial and axi-planar states at zero pressure and this might be interesting to investigate further.

\begin{figure}[t]
\centerline{\includegraphics[width=3.4in]{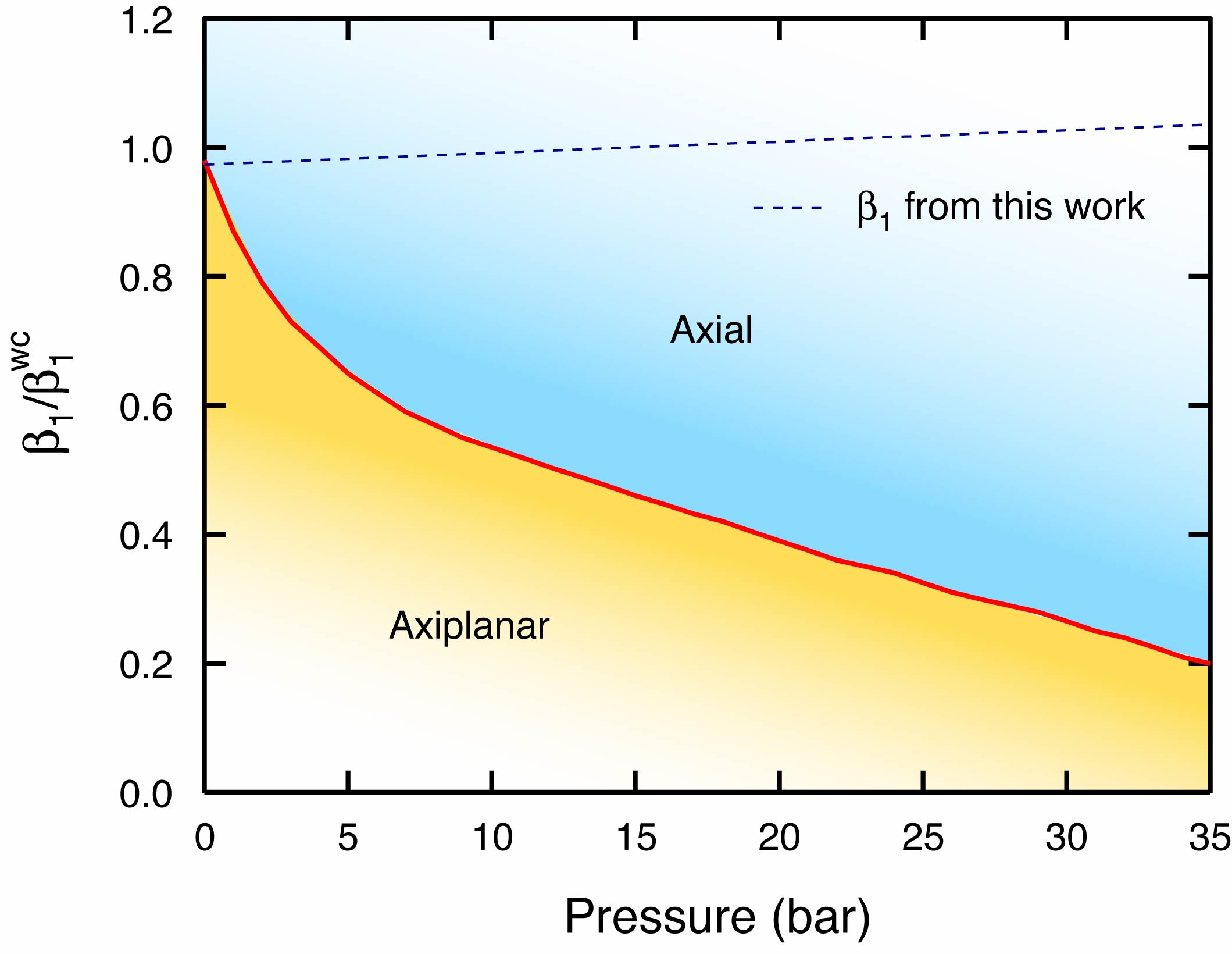}}
\begin{minipage}{0.93\hsize}
\caption {Phase diagram for axial and axi-planar states with
$\beta_{1}$ as a parameter. The choice of
$\beta_{1}$ with our model (dashed  line) places the $A$-phase in  the
region of the axial state.}
\end{minipage}
\end{figure}
\noindent

\subsection{The Robust Phase in Aerogel} Full determination of all five
$\beta_{i}$'s has important implications for superfluid $^{3}$He in
aerogel. Within the context of scattering  models we can calculate the
appropriate modifications to the $\beta_{i}$'s and explore the predicted
stability of various superfluid states. There are multiple different scattering
models for $^{3}$He in aerogel, e.g., the homogeneous  isotropic scattering
model  (HISM)\cite{Thu98} and inhomogeneous isotropic scattering models
(IISM)\cite{Thu98, Han03,Sau03}. We use the IISM of Sauls and Sharma\cite{Sau03}, a modification of 
the HISM of Thuneberg
{\it et al.}\cite{Thu98}. The $\beta_{i}$'s in the IISM are modified through:
\begin{eqnarray} \left(
\begin{array}{c}
\beta_{1}^{a}\\
\beta_{2}^{a}\\
\beta_{3}^{a}\\
\beta_{4}^{a}\\
\beta_{5}^{a}\\
\end{array}
\right) = \beta_{0}^{a}
\left(
\begin{array}{c} -1\\ 2\\ 2\\ 2\\ -2\\
\end{array}
\right) +b
\left(
\begin{array}{c} 0\\ 1\\ 0\\ 1\\ -1\\
\end{array}
\right) +
\left(
\begin{array}{c}
\Delta \beta_{1}^{sc,a}\\
\Delta \beta_{2}^{sc,a}\\
\Delta \beta_{3}^{sc,a}\\
\Delta \beta_{4}^{sc,a}\\
\Delta \beta_{5}^{sc,a}\\
\end{array}
\right),
\end{eqnarray}
\\*[-20pt]
\begin{equation}
\beta_{0}^{a}={N(0) \over {30 (\pi k_{B}T_{c})^{2}}}
\sum_{n=1} {1 \over {(2n-1+x)^{3}}},
\end{equation}
\\*[-20pt]
\begin{equation}
b={N(0)\over{9(\pi
k_{B}T_{c})^{2}}} \left( \mathrm{sin^{2}} \delta_{0} - {1 \over 2}\right)
\sum_{n=1} {x \over {(2n-1+x)^{4}}},
\end{equation}
where $x=\hat x/(1+\zeta_{a}^{2}/\hat x)$, $\zeta_{a}=\xi_{a}/\lambda$,
$\hat x=\hbar v_{F}/2 \pi k_{B}T \lambda$, $\xi_{a}$ is the strand-strand
correlation length, $\lambda$ is the transport mean free path for $^{3}$He
quasiparticles, and $\delta_{0}$ is the $s$-wave scattering phase shift.

With the five $\beta_{i}$'s for  bulk superfluid given in Table II, we
calculated the effects on the
$\beta_{i}$'s of scattering from the aerogel strands. We distinguish 
these coefficients
from bulk $^{3}$He with a superscript, $\beta_{i}^{a}$. We assumed 
unitary scattering,
$\delta_{0} =
\pi/2$ with
$\lambda$ = 150 nm and
$\xi_{a}$ = 40 nm.  These parameters are typical of 98\% porosity
aerogels\cite{Hal04}. The effects of scattering in the weak coupling 
approximation are
included in both $\beta_{0}^{a}$ and $b$. In addition, the 
$\beta_{i}^{a}$'s  will have a
strong coupling component that will be modified by elastic 
scattering.  We accomodate this
by  rescaling the $\Delta \beta_{i}^{sc}$'s with a factor
$T_{ca}$/$T_{c}$, since strong coupling effects\cite{Rai76} are linear in
$T_{c}/T_{F}$.  The results of the calculation,
$\beta_{i}^{a}/\beta_{0}^{a}$, are tabulated in the last five columns of
Table II.  For this choice of aerogel  parameters the superfluid state is
not stable below a pressure of 5  bar as reported by Matsumoto
{\it et al.}\cite{Mat97}, and hence the table  is blank below this pressure.

A direct consequence of the modification of $\beta_{i}^{a}$'s according to the
scattering model is the enhancement of  relative stability of the
$B$-phase with respect to the $A$-phase for $^{3}$He in aerogel. For
either the HISM or IISM, the isotropic state ($B$-phase) is found to be
stable over the entire  pressure range.  However, superfluid
$^{3}$He in aerogel has a metastable $A$-like phase that has been  clearly
observed\cite{Bar00,Ger02,Naz04} in various samples on cooling below
$T_{c}$. Although the exact nature of this phase  is still in question,
it is known that the metastable phase is an equal-spin-pairing
state\cite{Spr95}, similar  to the bulk $A$-phase, hence it is referred to
as an $A$-like phase. However, lack of understanding of the orbital part of the
order parameter makes the identity of the  state  less clear. Furthermore,
the question of stability of any equal-spin-pairing state with repect to
the aerogel $B$-phase  relies on an understanding of the appropriate
$\beta$-parameters for which we have no direct independent information.
Volovik\cite{Vol96} has argued that the axial state in the presence  of
quenched anisotropic disorder  cannot exist as a spatially homogeneous
superfluid owing to arguments from Imry and Ma\cite{Imr75}. If  the
metastable phase is in fact the axial state, the order parameter  would not
have long range orientational order, a state which Volovik has called  a
superfluid glass.  With a  different approach, Fomin\cite{Fom04} has
argued that there are other $p$-wave pairing states which are also
equal-spin-pairing but  do not suffer from the same difficulty, and that
these might be candidates for the  metastable aerogel phase.  Such phases
would be robust in the  presence of anisotropic scattering, meaning that
$A_{\mu i}A_{\mu  j}^{*} + A_{\mu j}A_{\mu i}^{*} \varpropto \delta_{ij}$
where $\delta_{ij}$ is the Kronecker delta.\cite{Fom04} NMR experiments have
been performed on
$^{3}$He in
$97.5\%$ aerogel  which support the view that the metastable $A$-like
phase  is in fact a robust state\cite{Ish06},  but other measurements
\cite{Cho04,Osh04,Dmi06} appear to be inconsistent with this
interpretation.   The free energy for the robust state\cite{Fom04} can be
expressed as,
\begin{equation} F_{R}=-\alpha^{2}/4\beta_{R},
\end{equation} where,
\begin{equation}
\beta_{R}=(\beta_{13}+9\beta_{2}+5\beta_{45})/9.
\end{equation}

\begin{figure}[t]
\centerline{\includegraphics[width=3.4in]{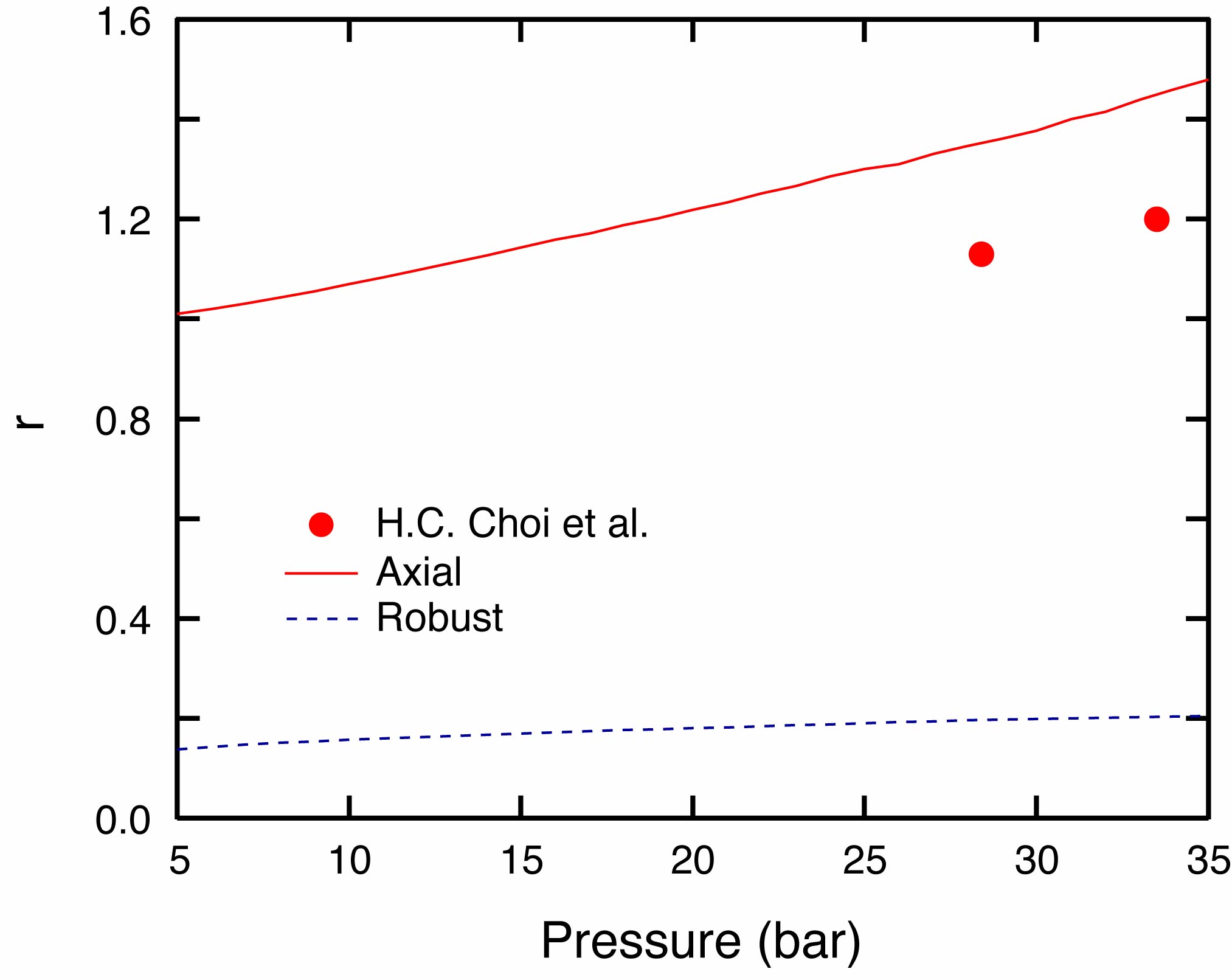}}
\begin{minipage}{0.93\hsize}
\caption {The asymmetry ratio, $r$, for the $A_{1}-A_{2}$ splitting was
calculated for the axial state (solid red line) and the robust phase
(dashed  line) in aerogel, where we  used the IISM of Sauls and
Sharma\cite{Sau03} with a transport mean free path for
$^{3}$He quasiparticles $\lambda$ = 150 nm and strand-strand correlation
length
$\xi_{a}$ = 40 nm, which match well to phase diagram measurements on the  same
sample by Gervais {\em et al}.\cite{Ger02}. The measurements of  H.C. Choi
{\em et al}.\cite{Cho04} (closed circles) are more consistent  with the
$A$-like phase of aerogel $^{3}$He being the axial state than the robust
state.}
\end{minipage}
\end{figure}
\noindent

Thermodynamic properties of the robust state have not been predicted
because it involves all five $\beta_{i}$'s beyond the four combinations
known to  us so far. However, the determination of
$\beta_{i}$'s from our model allows us to  investigate the properties of
the robust state.  First, we calculate the asymmetry ratio of the
$A_{1}$-$A_{2}$ splitting in aerogel. For the $A$-phase this ratio is
expressed in terms of the
$\beta_{i}$'s given by Eq. 12. In the case of the robust
state the ratio $r_{R}$ is given by\cite{Fom04,Cho04},
\begin{equation} r_{R} ={\beta_{15} \over
{\beta_{13}+9\beta_{2}+5\beta_{45}}}.
\end{equation}
With the values of the $\beta_{i}$'s from Table II, the
asymmetry  ratio $r_{R}$ is found to be $\sim 0.2$, considerably smaller
than what has been found experimentally\cite{Cho04}, $r_{R} \gtrsim 1.0$.
These results are compared in Fig. 6.

Second, we calculate the relative stability of the robust state with
respect to the $B$-phase over the pressure range from zero to 34  bar  with
$\beta_{1}$ for bulk $^{3}$He as a parameter subject to the
    constraints of the four experimentally known combinations of the
$\beta$'s given in Table I. These results are shown in Fig. 7.  For the
robust state of  $^{3}$He  in aerogel to be stable, $\beta_{1}$ would have
to be significantly different from the value derived from our model, 
assuming the form of
the free energy in Eq. 1.

\begin{figure}[t]
\centerline{\includegraphics[width=3.4in]{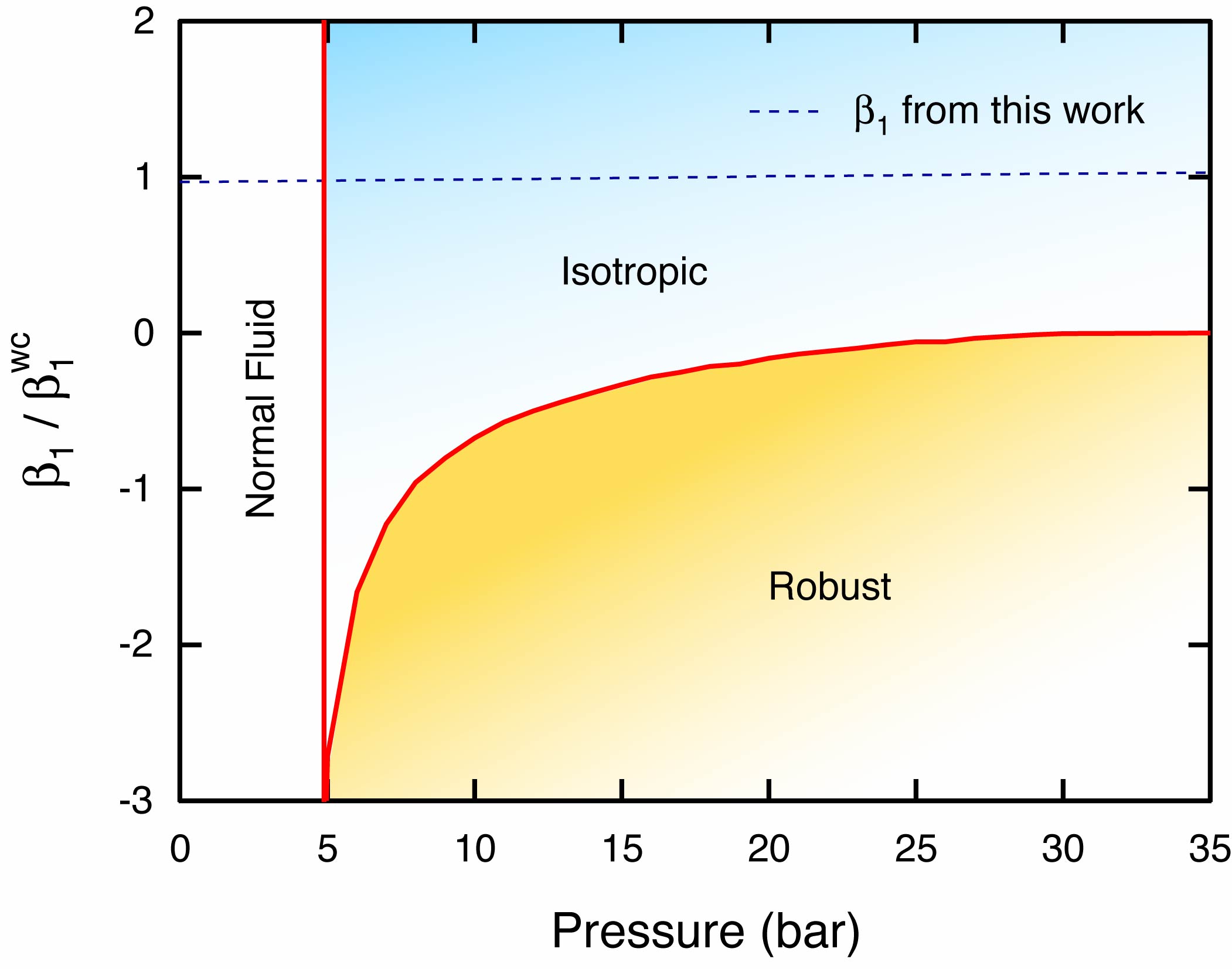}}
\begin{minipage}{0.93\hsize}
\caption {Phase diagram for the isotropic phase and the robust phase 
for  $^{3}$He in
98 \% porosity aerogel  with
$\beta_{1}$ (bulk) as a parameter. Our choice of $\beta_{1}$ (dashed   line)
makes the isotropic phase more stable than the robust phase.}
\end{minipage}
\end{figure}
\noindent

\section{Conclusions}

We have investigated the experimental basis for determining strong
coupling as a function of pressure
    for superfluid $^{3}$He based on analysis of our NMR  data.  Given
the limitation that we have only four experimentally identifiable
$\beta$-coefficient combinations, we developed a phenomenological  model
with two assumptions: 1) that superfluid $^{3}$He is predominantly
weak coupling at low pressure and 2) that the pressure dependence of
$\beta_{1}$ can be taken from Sauls and Serene's calculation. This model
provides us with all five
$\beta$-coefficients. Using this model we calculated the surface free
energy at the $A$-$B$ interface and compared with experiment. Although the
measurement does not have high enough resolution to validate our  model,
it is not in disagreement. The model is also consistent with the general
consensus that the so-called
$A$-phase is the axial phase rather than the axi-planar phase.   We used  our
values of the
$\beta_{i}$'s to calculate the corresponding strong coupling effects  for
superfluid $^{3}$He in aerogel. We find that the $B$-phase
    is stable at all pressures.  We compared  the
relative stability of the robust state proposed by Fomin with that of the
$B$-phase. The robust state is unstable relative to either the isotropic
$B$-like phase or the axial state. Furthermore, the asymmetry ratio,
$r_{R}$, of the $A_{1}$-$A_{2}$  splitting for superfluid $^{3}$He in
aerogel was calculated for the robust state and it was found to be
significantly smaller than the experimental values. Our interpretation is  that
the
$A$-like aerogel phase  is not a robust state based on the free 
energy expansion given in
Eq. 1.

\section{Acknowledgements}

We gratefully acknowledge discussions with Jim Sauls and support from the NSF
DMR-0244099.

\end{document}